\documentclass[pdflatex,sn-mathphys-num]{sn-jnl}

\usepackage{graphicx}%
\usepackage{multirow}%
\usepackage{amsmath,amssymb,amsfonts}%
\usepackage{amsthm}%
\usepackage{mathrsfs}%
\usepackage[title]{appendix}%
\usepackage{xcolor}%
\usepackage{textcomp}%
\usepackage{manyfoot}%
\usepackage{booktabs}%
\usepackage{algorithm}%
\usepackage{algorithmicx}%
\usepackage{algpseudocode}%
\usepackage{listings}%
\usepackage{tabularx}
\newcolumntype{C}[1]{>{\hsize=#1\hsize\raggedright\arraybackslash}X}
\usepackage{longtable}
\usepackage{xurl} 

\theoremstyle{thmstyleone}%
\theoremstyle{thmstyletwo}%

\theoremstyle{thmstylethree}%

\begin{document}

\title[Article Title]{From Guidelines to Practice: Defining Requirements for AI Red-Teaming Tools}





\author*[1]{\fnm{Gia Phong} \sur{Tran}}
\author[1]{\fnm{Thai Duong} \sur{Huynh}}
\author[1]{\fnm{Randolph} \sur{Loh}}
\author[1]{\fnm{Balachandar} \sur{Gowrisankar}}
\author[1]{\fnm{Zihao} \sur{Wang}}
\author[1]{\fnm{Moe Thu} \sur{Kyaw}}

\author[2]{\fnm{Jun} \sur{Sun}}
\author[3]{\fnm{Adams Wai-Kin} \sur{Kong}}
\author[1]{\fnm{Vrizlynn L. L.} \sur{Thing}}




\affil[1]{\orgdiv{Cyber Security Strategic Technology Centre}, \orgname{ ST Engineering}, \orgaddress{\country{Singapore}}}

\affil[2]{\orgdiv{School of Computing and Information Systems}, \orgname{Singapore Management University}, \orgaddress{\country{Singapore}}}

\affil[3]{\orgdiv{College of Computing and Data Science}, \orgname{Nanyang Technological University}, \orgaddress{\country{Singapore}}}


\abstract{Artificial Intelligence (AI) has seen significant developments and increased usage 
in recent years. AI benefits various tasks, automates decision-making, and enhances 
efficiency. However, AI also introduces additional attack surfaces for threat actors 
to perform attacks like prompt injection and jailbreaking, making security a great 
concern. AI security testing aims to imitate adversarial attacks or malicious inputs 
via manual testing or with the assistance of tools. Thus, it is critical that 
organizations are aware of the capabilities and limitations of the tools before using 
them. Existing studies provide insights on the overall architecture of AI security 
testing tools and raise concerns on their reliability, but their analysis is limited 
to attack performance. Selection of a tool for AI risk assessments should also depend 
on their alignment with the recommendations of industry standards, but to our 
knowledge, previous studies have not utilized such guidelines. To bridge this gap, we 
analyze several industry standards and guidelines to identify the requirements of AI 
testing tools. We discuss and highlight desirable features across various AI security 
testing tools for a variety of different AI systems. We evaluated six tools across six 
Large Language Models, reported on our insights, and identified gaps that should be 
considered by future tools.}




\maketitle

\section{Introduction}

Artificial Intelligence (AI) is an area of computer science that aims to simulate human learning and enable problem-solving capabilities in computers. AI techniques capture relationships in the data and utilize them to anticipate results or generate synthetic samples \cite{1}. AI systems comprise complex architectures and require immense computational power to facilitate training and inference decisions. AI enhances productivity by automating tasks in fields not limited to healthcare, finance, and transportation \cite{2}.

Generative AI (GenAI) is a sub-category of AI and has seen tremendous growth in recent years. For one, the widespread application and business opportunities of large language models (LLMs) promoted the proliferation and development of AI. LLMs had evolved from earlier task-specific language models to versatile multi-tasking models, which can now be fine-tuned with minimal data learning \cite{3}. LLMs have exhibited contextual understanding of natural language and can produce text that resembles human writing and execute instructions, potentially adaptable to different industries and environments \cite{4}. They are efficient in text generation, summarization, translation \cite{5,6,7}, and can be effectively applied in a wider range of application scenarios such as code generation, research and education in medicine, automated financial report generation, and forecasting market trends and analysis of investor sentiment \cite{8,9,10}. LLMs were shown to be comparable to or even outperform humans in tasks like summarizing clinical text \cite{11} and achieving a top 10\% score on a simulated exam in academic benchmarks \cite{12}.

LLMs have brought revolutionary changes to workflows in many fields. ChatGPT \cite{55} alone has over 400 million weekly users \cite{13}. Meta AI's LLaMA \cite{14} and Google's PaLM \cite{15} are also popular. Forbes predicts that the AI market size is expected to reach USD 1,339 billion by 2030 while an IBM report mentioned that about 42\% of enterprise-scale companies have already deployed AI in their businesses \cite{16,17}. AI will no doubt accelerate operations, development, and innovation.

However, AI does not come without flaw, and AI related cyberattacks were reported. Hugging Face, a popular repository for AI resources, was found to contain poisoned models that may potentially grant privileged access to a victim’s machine through code manipulation and modification \cite{19}. Microsoft uncovered a jailbreak vulnerability that can extract personally identifiable information (PII) and harmful content from GenAI models \cite{20}. The attack simply involves sending a prompt that tells the model to augment its encoded security features. Researchers from HiddenLayer discovered that the Gemini Pro model is susceptible to the inverse scaling property of LLMs \cite{21}. They constructed malicious prompts from synonyms to extract the system prompt, which should only be accessible to model administrators.

There are plenty of tools in the market to help organizations evaluate the security and safety of AI applications. AgentFence \cite{61}, Garak \cite{33}, Giskard \cite{58}, Promptfoo \cite{59}, Prompt Fuzzer \cite{60}, and Whistleblower \cite{62} are some examples. With a plethora of such tools in the market, a common question arises: How do we select the right tool for our use case with specific requirements and restrictions? A study performed by Brokman et al. \cite{63} provides insights on the overall architecture of popular AI security testing tools and their inconsistencies in attack evaluation. While their study provides an initial foundation for comparing tools, their analysis does not shed light on whether the tools can satisfy the needs and requirements of organizational AI risk management. It is necessary to recognize the requirements for security, governance, and usability to assess a tool’s suitability for AI risk management and propose directions to fill missing gaps in their architecture.

Standards bodies and global entities have released materials, frameworks, principles, and guidelines \cite{18, 22,23,24,25,26,27,28,29,30,31,32} to help organizations control and minimize AI risks. They provide the understanding to establish effective strategies for secure development, deployment, and engagement of AI systems. They contain expert recommendations regarding the desirable features of AI security testing tools and thereby evaluate their capabilities for performing AI risk assessments.

In this paper, we performed an in-depth analysis of various guidelines to understand their strategies to securely develop and deploy AI systems, identified desirable features for AI security testing. We aim to provide awareness to organizations when adopting AI security testing tools with a set of proposed desired features. Here are the main contributions of this work:
\begin{itemize}
    \item We analyzed various industry standards and guidelines to understand their strategies and the role of security testing tools for AI risk management. 
    \item We highlighted desirable features of AI security testing tools from our analysis. 
    \item We examined how desirable features are reflected in practice by reviewing frameworks and guidelines on top of experimenting, comparing, and analyzing several open‑source LLM red‑teaming tools. 
\end{itemize}

This is the structure of the remaining sections: Section II details background information about relevant guidelines for securing AI systems and an overview of different AI security testing tools. In Section III analyzed how AI security testing tools can contribute to fulfilling recommendations of AI security guidelines and identified a list of desirable features. We determined how the desired features relate to existing guidelines in Section IV. Section V describes the experiments and results shared in Section VI. Conclusion is in Section VII.

\section{Background}

\subsection{AI security frameworks and guidelines}

In face of the inherent challenges and risks of AI systems and models, standards bodies and global entities have released documents and guidelines \cite{23,24,25,26,27,28, 30, 38, 39} to address a broad spectrum of security use cases across the AI System Development Life Cycle (AI SDLC). Organizations can then better understand security concerns of AI systems or adopt them into their own security frameworks and strategies for the secure development, deployment, and engagement of AI systems while managing and mitigating the associated risks.

\subsubsection{Core frameworks and guidelines}

Guidelines and frameworks developed by the National Institute of Standards and Technology (NIST), The MITRE Corporation, and Open Worldwide Application Security Project (OWASP) are widely referenced and adopted into the security guidelines and frameworks of many organizations. 

NIST developed many cybersecurity standards, guidelines, best practices, and other resources to meet the needs of industries, federal agencies, and the broader public like the Cybersecurity Framework 2.0 (CSF) \cite{65}. Specific to AI, the NIST AI Risk Management Framework (AI RMF) \cite{22} helps organizations identify risks presented by AI systems and provides recommendations to address them across four core functions: Govern (policies, culture, and oversight), Map (risk identification), Measure (risk analysis and tracking), and Manage (risk prioritization and mitigation). In addition, the NIST Trustworthy and Responsible AI report \cite{29} provides a taxonomy of concepts, definitions, and terminology encompassing key Machine Learning (ML) methods and details Adversarial ML (AML). Organizations and entities may refer to the taxonomy and terminology of attacks, encompassing both Predictive AI (PredAI) and Generative AI (GenAI), and mitigation strategies of AML.

Similarly, OWASP aims to enhance the security of software and relevant applications through various community-led open-source projects with many of them relevant to AI systems. The OWASP Machine Learning Security Top Ten \cite{66} lists the top ten security issues of ML systems and the OWASP Top 10 for Large Language Model Applications \cite{57} identifies critical security vulnerabilities of LLM in real-world applications. The OWASP AI Exchange \cite{67} is a community-led compilation of state-of-the-art topics and global discussion on the security and privacy of all types of AI and data-centric systems. An extensive framework details threats, controls, and best practices for secure, private AI usage. It encompasses topics on the principles and requirements of different international legislations, security, and privacy standards and how they would apply to AI systems. The OWASP AI Exchange helped develop the EU AI Act \cite{68}, ISO/IEC DIS 27090 \cite{69}, and 27091 standards \cite{70}. The OWASP GenAI Security Project \cite{71} is also a community-driven, open-source project that directs stakeholders to resources to better identify, mitigate, and document security and safety concerns of GenAI applications and systems. 

Lastly, the National Cybersecurity Federally Funded Research and Development Centers, operated by The MITRE Corporation, investigates cybersecurity challenges surrounding AI, IoT, Zero Trust, and Space to provide secure technologies and to improve the cybersecurity posture and protect sensitive data and systems across industry and government. A key contribution being the MITRE ATT\&CK resource center that details real-world adversary tactics and techniques \cite{73}. The MITRE Adversarial Threat Landscape for Artificial-Intelligence Systems (ATLAS) is a knowledge repository for AI systems \cite{56}. ATLAS informs stakeholders of realistic threats that real-world adversaries pose to AI systems.

\subsubsection{Global, international entities, and consortiums}

Global and international entities came together for dialogue and established directives and approaches to address concerns regarding the secure AI systems development and usage. They provide consensus and influence across countries and regions. 

The Hiroshima AI Process (HAIP) \cite{75} launched in May 2023 at the G7 Hiroshima Summit. The HAIP Comprehensive Policy Framework came into fruition from the discussions of the G7 Digital \& Tech ministers. Endorsed by the G7 Leaders, the HAIP Comprehensive Policy Framework is expected to support a diverse range of actors, expanding multiple governments, sectors, academia, and society. Key resources of the framework include the Hiroshima Process International Guiding Principles and Code of Conduct for Organizations Developing Advanced AI System and Guiding Principle for All AI Actors \cite{30,31,32}. The global governance initiative emphasizes innovating opportunities in AI systems while mitigating risks and challenges. It aims to boost safe and secure AI, citing 11 principles for the design, development, deployment, provision, and use of AI systems.

The Organization for Economic Co-operation and Development (OECD) AI Principles \cite{74} offer a base for international cooperation and interoperability that policymakers can adopt for the development and deployment of innovative and trustworthy AI. Five values-based principles that advocate innovative and trustworthy AI, honoring human rights and democratic values. There are also five inter-complementing recommendations for policymakers to promote long-term public and private investments for interdisciplinary research and development for an inclusive, open, and trustworthy digital ecosystems with interoperable governance, policies, regulatory frameworks, and assessment mechanisms to prepare and transform international markets through skill and competence building, collaboratively with consensus-driven initiatives across governments and stakeholders. The OECD also launched a global framework to monitor the application of the HAIP Code of Conduct \cite{32}.

National level security agencies also jointly produced cybersecurity information and reports to provide guidance for the design, development, and use of AI systems. These agencies are not limited to the National Cyber Security Centre (NCSC) of the United Kingdom, the Australian Signals Directorate (ASD), the United States' National Security Agency's Artificial Intelligence Security Center (NSA AISC). Guidelines for secure AI system development \cite{27} details four key focus areas of the AI SDLC: secure design, secure development, secure deployment, and secure operation and maintenance. It emphasizes the concept of secure by default, stresses taking responsibility for customer security, being open and accountable, and organizing teams and leadership so that security is treated as a top business priority. The Engaging with Artificial Intelligence publication \cite{28}, however, focuses on the use of AI systems. It compiles threats to AI systems and mitigations for organizations to address risks while interacting with these systems. The Deploying AI Systems Securely report \cite{26} shares best practices for deploying secure and robust AI systems. It aims to strengthen the security of AI systems and environments with proper management of known vulnerabilities and weaknesses, leverage methodologies, and controls to address malicious activities that target AI systems and relevant data services for secure deployment, operation, and maintenance of AI systems. The AI Data Security information sheet \cite{76} provides an overview of data security for the development, testing, and operation of AI systems throughout the AI SDLC. It describes best practices to secure data, data supply chain risks, maliciously modified data, and data drift and corresponding mitigation strategies. It focuses on principles to protect sensitive, proprietary, or mission critical data and to provide a solid foundation for the security of AI data and ensure the reliability of AI systems.

The European Union Agency for Cybersecurity (ENISA) promotes cyber resilience, safeguards communication and data, and keeps online society and economy secure. ENISA has produced the framework for good AI cybersecurity practices (FAICP) \cite{77} that advises national competent authorities, stakeholders, and communities on steps to secure AI systems, operations, and processes through known best practices. It is segmented into three layers. The cybersecurity foundations layer details basic cybersecurity knowledge and practices for Information Communication Technologies (ICT) environments that host AI systems. The AI-specific cybersecurity layer details cybersecurity practices applicable throughout the AI systems, components, and SDLC. The sector-specific cybersecurity for AI layer details best practices to secure AI systems from sector specific threats. ENISA is additionally supported by the EU's Cybersecurity Act \cite{78} which has established the cybersecurity certification framework \cite{79} for ICT products and services referencing standards and methodologies to evaluate the security of ICT assets hosting AI components. 

The EU AI Act \cite{68} is the European regulation on AI. It assigns risk categories to AI systems. Prohibited AI systems have unacceptable risks that can negatively influence decision making, infer sensitive information, and heavily impact security, safety, and privacy. Certain systems that are employed in use cases such as biometrics and critical infrastructure are classified as high-risk, whose providers are required to have proper management of data and risks, and to ensure the quality and performance of the AI system. It inclines undergoing third-party compliance assessments. There are also General Purpose AI (GPAI) models and systems. GPAI models are trained with extensive amounts of data to perform generic tasks, which can be used directly or integrated into other AI systems to serve a variety of purposes. GPAI providers need to provide sufficient informational and technical documents detailing the training and testing process with evaluation results detailing their capabilities and limitations. Cybersecurity protection must be established, evaluated, and assessed by the ability to mitigate possible systemic risks of GPAI with tracking, documentation, and reporting capabilities. The Guidelines for GPAI Models \cite{80} details information on the definition and scope of GPAI models, obligations over the GPAI model lifecycle, significance resulting from systemic risk, and communication responsibilities of providers. The Code of Practice \cite{81} provides a framework for developers of GPAI to adjust to the EU AI Act. Specifically in the areas to promote Transparency, ensure alignment to Copyright laws, and factors for the Safety and Security of the GPAI models and systems. 

The ASEAN Guide on AI Governance and Ethics \cite{82} can be adopted into organizational and governmental agency practices to design, develop, and deploy AI systems in a responsible manner. It describes guidelines, provides recommendations, and details seven guiding principles for the AI Governance Framework to promote care for AI technologies and systems. It recommends that AI systems be transparent and explainable for users to understand and be confident in decisions and recommendations put forth by AI systems and to make informed decisions with sufficient documents from developers and deployers. AI systems should also be fair and equal such that algorithmic decisions do not result in unjust impacts, unfair biasness, or discrimination. Mechanisms are in place to identify, evaluate, and address any risks for the safety and security of AI system, including attacks performed by malicious adversaries, and for humans to mediate when needed. In addition, mechanisms that ensure the privacy and governance of data throughout the AI system lifecycle. It also recommends periodical reviews of data protection laws, data governance policies and legislation, and ethical principles as appropriate to accommodate the wellbeing and benefit of human society. It asks developers and deployers to be accountable and act with integrity, for AI systems to function properly and are compliant. AI systems should be robust, reliable, and perform consistently even under complex and unknown conditions. The Guide's AI Governance Framework shares key areas that organizations should consider for the responsible use of AI. It suggests incorporation of existing or new internal governance structures and measures to guide the design, development, and deployment of AI systems, assess risks associated with AI-augmented decision-making functions and determine the appropriate level of human involvement, manage underlying tasks associated with AI system operations, and establish appropriate channels to manage communications, interactions, and relationships with stakeholders. Lastly, the Guide considers national and regional level recommendations for policymakers in ASEAN and encourages the alignment and interoperability of AI frameworks across jurisdictions within the region.

The Cyber Security Agency of Singapore (CSA) developed Guidelines on Securing AI Systems \cite{18} for the secured use of AI systems throughout their lifecycle. It helps identify potential security risks and provide mitigating strategies that can be applied throughout the five key stages of the AI SDLC: planning and design, development, deployment, operations and maintenance, and end of life. A Companion Guide on Securing AI Systems \cite{24} was also developed alongside as a community-driven resource, with practical security control measures to support system owners to secure their AI systems. 

Safety and security of AI systems are of global interest. There is great significance in testing tools that assist AI developers, systems owners, and stakeholders to ensure secure development and usage of AI systems.

\subsection{AI systems and associated attacks}

AI systems can be categorized into two main branches \cite{29}. One, PredAI that comprises models like logistic regression, support vector machines (SVM), and convolutional neural networks (CNN), learn decision boundaries to differentiate between classes. Second, GenAI comprises models like generative adversarial networks (GANs), generative pretrained transformers, diffusion models, and multi-model content generation, learning to generate diverse types of content and media. It is important to note that attacks may overlap and apply across both branches of AI systems. 

\subsubsection{Adversary goals and capabilities}

Attacks aim to compromise the availability, integrity, and privacy of AI systems \cite{29}. Attacks on availability affect robustness and performance, can include poisoning attacks in which adversaries have some level of control over the training dataset or model. Attacks on integrity affect the outputs of AI systems and models, causing them to react incorrectly. They include evasions and poisoning attacks in which adversaries create adversarial examples to induce misclassification. Attacks on privacy can leak sensitive information about the training data or model such that adversaries may be able to reconstruct, infer data used to train a model, or learn properties, architecture, or parameters regarding a model. Adversaries may also seek to abuse and repurpose AI systems for their own agenda in a way that violates their original intended use. 

Adversaries exhibit various capabilities at different phases of AI systems and models. During the training phase, adversaries may seek to control a subset of the training data or the model. Therefore, degrading the availability and integrity of AI systems and models. Adversaries can poison training data by inserting their own or modifying it and influence model parameters by embedding malicious functions in the AI system. In more realistic scenarios, adversaries may have limited access to some labels or not be able to directly manipulate the labels used to train AI systems and models as demonstrated in clean-label poisoning attacks. 

During the testing and deployment phase, adversaries may seek to control testing data by adding perturbations to testing samples to generate adversarial examples when gaining access to queries submitted to AI systems and models. Adversaries can use adversarial examples to incite behaviors that compromise availability, integrity, and privacy. They can induce the AI system or model to misclassify or activate backdoors. Adversaries may seek to take control of the resources and source code that are passed to AI systems and models to influence their behavior. Access to and knowledge of AI systems and models also influences the adversaries' capabilities to conduct attacks in a white-, black- or gray-box setting. 

\subsubsection{Attacks on Predictive AI}

PredAI systems can be modeled as a generalized data processing pipeline that follows a primitive sequence of operations across the training and development, testing and deployment phases \cite{29}. In both phases, the AI system receives input data from a source, input data is preprocessed for the AI system, and AI model receives the processed input data. During the final output, a trained AI model is produced in the training and development phase, while the AI model produces actionable outputs in the testing and deployment phase. Numerous attacks can occur at different operation sequences across both phases and negatively impact the resulting AI system. The AI system is susceptible to poisoning attacks during training time while being susceptible to evasion and privacy attacks during testing and deployment time. 

Poisoning attacks are when adversaries interfere during training time to poison data or the model. They affect the availability and integrity of AI systems and models, including input manipulation, availability poisoning, targeted poisoning, and backdoor poisoning attacks. 

Evasion attacks are the most common type of attack. where adversaries provide malicious inputs or adversarial examples to trick the AI system to misinterpret or misclassify the input data by some arbitrary class of choice. In the white box setting, adversaries aim to minimize perturbations and support multiple distance metrics to optimize adversarial example generation. In the black box setting, adversaries can only query the AI system on various data samples with information about settings or past inputs. In a more restrictive setting with less access to information, decision-based attacks only use the final predicted labels of the model. Transferable attacks are when adversarial examples generated for a substitute model can be transferred to a target model given some intersecting decision boundaries for adversarial subspaces with higher dimensionality.

Privacy attacks try to gain as much knowledge as possible about the AI system and model in a way that violates privacy of the information being learned. This ranges from information pertaining to the data used in training to information regarding the AI system and the models themselves. Privacy attacks include membership inference, data reconstruction, property inference, and model extraction attacks.

\subsubsection{Attacks on Generative AI}

In GenAI, challenges associated with the security and privacy of LLMs have become increasingly dire. Alongside associated vulnerabilities and challenges that encompass attacks on AI and non-AI inherent model vulnerabilities, and on the different modules of LLMs. Das et al. reviewed the security and privacy challenges across nine representative attack and defense mechanisms on LLMs \cite{54}. On top of attacks on PredAI systems, the NIST AI 100-2 E2025 report \cite{29} gives attention to additional categories of attacks relevant to GenAI systems. Where attacks can be linked to more multiple attack classes to achieve different attacker goals. 

Supply chain attacks take note of the inherited software supply chain risks and exploits artefacts relevant to specific data-based statistical properties of AI systems. They include poisoning attacks that try to insert backdoors into pre-trained models. This assumes attackers with more access to knowledge, and control of their targeted AI system and possess the necessary resources to execute and accomplish the attack. For example, Yan et al. devised a Virtual Prompt Injection backdoor attack that allows the model to make a designed response in specific departure scenarios while behaving normally in other scenarios \cite{84}. Attackers can guide a model without any explicit injection of the model's inputs. 

Direct prompting attacks mainly focus on the input and output layers of GenAI models and do not normally target the internal structure. They do not involve internal modification of the model, require lower hardware demands, and are widely implemented by GenAI testing tools. For example, the DAN (Do Anything Now) \cite{83} attack module from the GenAI offensive testing tool Garak is a type of prompt injection technique to bypass the ethical constraints of GenAI models in black-box settings. The attack tries to convince the GenAI model to ignore the preset restrictions for unrestricted answers with tweaked prompts. 

Indirect prompt injection attacks assume third party attackers can manipulate system relevant resources without directly interacting with the GenAI system. Attacks on availability can disrupt the legitimate use of GenAI system. Attacks on integrity can influence GenAI systems to subtly deviate from its usual behavior, or to maliciously perform attacker-specific tasks, becoming untrustworthy. Attacks on privacy can lead to indirect leaking information of interconnected systems and can persuade end-users to share potentially sensitive information.  

Attacks that require an attacker to have a deep understanding of the model's mechanisms complicate their integration into existing GenAI testing tools. Therefore, most attack tools tend to focus on attack techniques with broader applicability like direct prompting attacks.

\subsection{AI security testing tools}
In response to the ever-evolving threat landscape and continuous effort that explores the security and attacks on AI \cite{29, 53, 54}, AI security testing tools are further developed and updated. AI security testing tools are designed to detect weaknesses in AI systems by simulating adversarial attacks and tend to focus on either PreAI or GenAI models \cite{24}. PredAI security testing tools provide AI system and model developers with comprehensive attack libraries that include tests for evasion, poisoning, and membership inference attacks \cite{35,36,37,38, 49}. Tools may also cater to AI systems and models across a range of data modalities including image, tabular, audio, and video. GenAI, on the other hand, centers around generative models that learn to generate diverse types of content and comprises generative adversarial networks, generative pretrained transformers, and diffusion models. GenAI security testing tools encompass attacks like prompt injection, jailbreaking, and data leakage \cite{33, 34, 48}. AI systems and models are probed with carefully crafted prompts to elicit malicious or harmful responses. 

\subsection{Related works}
To the best of our knowledge, Brokman et al. presents the only work that focuses specifically on the evaluation of AI security‑testing tools \cite{63}. Their study provides an objective comparison of popular open‑source AI red‑teaming scanners, assessing both the diversity of their attack methods and their overall effectiveness. In contrast, this work takes a unique perspective by analyzing existing guidelines and frameworks to identify the necessary characteristics of AI security‑testing tools. This forms a more comprehensive and evidence‑based understanding of what such tools should encompass outlined in existing guidelines and frameworks. Organizations can consider and adopt these desirable features to determine which AI security testing tool best suits their use case. They include attention to the inclusion of comprehensive reports that detail the tests that were conducted, any vulnerabilities that were discovered, and respective mitigation strategies. 
\section{Guidelines analysis}

In this section, we analyzed seven guidelines and their recommendations to map how AI security testing tools can contribute to realizing some of these recommendations. We also identified several desirable features that should be examined in AI security testing tools from our analysis.
\subsection{Map applicability of AI security testing tools to guidelines}
Several frameworks and guidelines shared in Section II refer to security considerations for each stage of the AI SDLC. CSA’s Companion Guide on Securing AI Systems \cite{24} and NCSC’s Guidelines for secure AI system development \cite{27} describe the four main stages of the AI SDLC. In planning and design stage, AI security testing tools can be leveraged to help understand any threats, risks, and potential impacts associated with the design of the AI system. In the development stage, the focus of threats, risks, and potential impacts shifts from the design to the supply chain, AI related assets, development environment, and even technical debt. In the deployment stage, there is emphasis on the continuous integration/continuous delivery (CI/CD) pipeline. Here, AI security testing tools should support the efforts in securing the deployment infrastructure of AI systems, its continuous management, and the responsible release of AI systems. These include security checks and evaluations before the release of AI systems and components, to assist end users in performing self-assessments, and to provide advice on the appropriate use of the AI system and components. In the operation and maintenance stage, AI security testing tools do not actively support monitoring AI systems inputs, outputs, and behavior. However, they can be used to help identify potentially malicious inputs, unsafe outputs, and behaviors that were identified in the previous stages of the AI SDLC, such as automated generation of additional guardrails to mitigate known successful attacks. However, newly identified vulnerabilities can be integrated into AI security testing tools for future assessments. CSA's Companion Guide on Securing AI Systems \cite{24} also describes a fifth stage to ensure measures are in place for the disposal and destruction of AI related components such as data and models. CSA’s Guidelines on Securing AI Systems \cite{18} also detailed measures and controls relevant to each stage of the AI SDLC. 

ASD’s Engaging with Artificial Intelligence \cite{28} provides guidance on engaging with AI systems, highlighting important threats and steps for managing risks. Whereby AI security testing tools can be applied across mitigation considerations for organizations. AI security testing tools can be used to assess the effectiveness of cybersecurity mitigations implemented by the organizations relevant to its jurisdiction to ensure that their AI systems are properly secure. Logging and monitoring mechanisms can also be assessed. Organizations can also evaluate incident response plans for business continuity of their AI systems. The tools can also help with the trial of an AI system, to help organizations learn about the AI system, and to test the limits and constraints in a separate environment before its integration. Apply the secure-by-design principle across the AI supply chain, identify and assess the threats and risk of components and assets associated with the AI system. 

The NSA AISC’s joint Cybersecurity Information Sheet Deploying AI Systems Securely \cite{26} describes best practices and controls to deploy AI systems securely. They include securing deployment environments, continuously protecting the AI system, and securely operating and maintaining AI systems. Deployment environments need to be secured through governance with the relevant security standards and requirements to promote collaboration. Security controls are in place to ensure that the deployment environments, both architectures and networks, are properly protected, hardened, and robust. AI security testing tools can assess and evaluate deployment environments accordingly. Assessments and evaluations can be conducted throughout the AI SDLC for continuous protection of AI systems. They validate AI systems, AI supply chain, AI artifacts, and assets such as testing protection mechanisms that secure exposed APIs, actively monitor AI system, and model behavior, and protect model weights. AI security testing tools can also be applied during operation and maintenance by both internal and external entities, to assess potential vulnerabilities and weaknesses. Such assessments can be in the form of audits and penetration tests. Assessment results can prompt stakeholders to take any necessary actions and respond in a timely manner. For example, pushing updates and patches regularly. The Hiroshima Process International Guiding Principles for Organizations Developing Advanced AI System \cite{30} describes eleven guiding principles for the design, development, deployment, provision, and use of AI systems. On top of the AI SDLC, AI security testing tools should also foster collaborative efforts to secure the AI ecosystem. It is important to publicly share information such as technical reports, instructions of use, and any relevant documentation or materials in a responsible, transparent, and meaningful manner. Relevant stakeholders can gain insights to perform any action needed to properly secure their AI systems and solutions. This encourages efforts to secure the AI ecosystem for global benefit, including the development of secure AI systems and solutions, AI security testing tools, and even international technical standards. 

The Infocomm Media Development Authority (IMDA) of Singapore proposed two Model Governance AI Frameworks. Although the frameworks are namely for AI governance, they include some principles that are also applicable to AI security testing. The Model AI Governance Framework \cite{23} contains eleven AI ethical principles that can be incorporated for the assessment of AI models. The AI Verify \cite{38} toolkit helps streamline the testing process to ensure user awareness for informed decision making. It also assesses the operations and results of AI systems and solutions. Consequently, the Model AI Governance Framework for Generative AI \cite{25} proposes nine dimensions for GenAI for the construction of comprehensive trustworthy AI systems. The frameworks complement the AI ecosystem to foster trusted development, and are aligned with internationally recognized frameworks, such as the EU AI Act \cite{68} and OECD \cite{39}. The AI ethical principles and dimensions provide various aspects to consider, test, and evaluate by leveraging AI security testing tools.

\subsection{Desirable features of AI security testing tools}

The technical documents and guidelines described in Section II-A share some similarities and differences in recommendations and views on the security of AI systems and components across the AI SDLC. We extracted statements from the guidelines such as "Red teaming should be an open and transparent process" and other implied ideas like comparing the integration capabilities of tools. We then grouped similar evidence into eight categories, which we identified as the eight desired features of AI security testing tools:

\begin{itemize}
    \item \textbf{Accessibility and Transparency:} This feature considers whether AI security testing tools are openly accessible and transparent. Potential adopters can inspect the tools to learn of their inner workings before integrating them into part of an organization's security framework. While this is straightforward for open-source distributions where source code is freely available, users of proprietary tools must rely on information provided by the developers or conduct trials to evaluate the capabilities and usefulness of the tool. NIST RMF \cite{22} advises the red teaming methodology to be an "open and transparent" process. Many guidelines comparing popular red teaming tools also evaluate the tool’s licensing approach \cite{24, 85, 86}. However, transparency can be a double-edged sword: while adopters can examine the tools to assess their trustworthiness, attackers can do the same to identify weaknesses and bypass detection.
    \item \textbf{Standardization:} This feature considers how testing tools align with security best practices and RMFs. Many frameworks \cite{22, 24} and regulations \cite{68} recommend or stipulate methodologies to secure and manage an AI system, and this feature considers how well the tool can help organizations satisfy such requirements. Some other frameworks \cite{56, 57} instead provide a formal taxonomy of definitions, and this feature also evaluates the tool’s ability to utilize them. Following standardization helps organizations objectively assess their security posture while satisfying mandatory compliance requirements. For example, CSA's Companion Guide on Securing AI Systems \cite{24} recommends that the risk assessment be conducted in accordance with applicable industry standards; IMDA's Model AI Governance Framework for Generative AI \cite{25} encourages organizations to follow common benchmarks and methodologies in a systematic manner to provide more consistent and comparable results; OWASP's GenAI Red Teaming Guide \cite{86} considers benchmarking against industry standards to be good practice. However, different organizations have different security postures, hence the organization should also consider the scope and coverage of the testing tools that are relevant to its security policies. 
    \item \textbf{Evidence Gathering and Documentation:} This feature evaluates the tool’s efficacy in gathering evidence and generating documentation. More specifically, we assess the assessment report’s level of detail (e.g., timestamps, prompt-response pairs) and its organization. A comprehensive report is essential for effective stakeholder communication \cite{24}. Furthermore, providing evidence throughout every stage of development and deployment lifecycle is a mandatory requirement for regulatory compliance \cite{68, 87}. 
    \item \textbf{Automation and Integration:} This feature considers the tool's ability to automate assessments of the AI system and model throughout the AI SDLC as required by the organization. It also considers the tool’s integration capabilities, since effective automation depends on how well the tool can operate within the existing infrastructure. The ability to automate assessments lets organizations delegate resources to other aspects of the AI SDLC. More specifically, as assessments should be conducted on a regular basis \cite{24, 26, 86, 88}, automation can significantly reduce the human resources needed and the risk of human error inherent in manual processes, enabling organizations to scale the operation efficiently. 
    \item \textbf{Models, Modality, and Threat Coverage:} This feature considers the types of AI systems, models (generative or predictive), data (such as images or texts), and risks that can be assessed by the testing tool. Many testing tools specialize in particular model types, with a lack of tools evaluating multi-modal models \cite{25}. However, it is still good practice to perform multi-modal vulnerability analysis \cite{86}. Even within the supported modality, tools should also maintain diverse and up-to-date test cases covering different threats \cite{86}, such as MITRE ATLAS \cite{56} or OWASP Top 10 for LLM \cite{57} in LLM case. Regardless, limited scenario coverage is not necessarily an indicator of poor quality, as providers may deliberately prioritize the scenarios most relevant to their specific use cases. Organizations should therefore select testing tools that are relevant to their requirements \cite{89}. 
    \item \textbf{Assessment Environments:} This feature considers environments in which the testing tool performs assessments. More specifically, organizations should consider whether the tool supports on-premises deployment or can only be hosted remotely. In many cases, on-premises deployment model entails higher prices except for open-source tools; however, they might be the only viable option for organizations with strict data governance policies. Moreover, different deployment models allow different testing approaches; remotely hosted tools typically only support black-box testing, whereas on-premises deployments may also permit white-box or grey-box testing. Black-box testing is more suitable for organizations with strict regulatory constraints and offers a more accurate simulation of an attacker’s perspective, while white-box testing allows greater visibility and enables more effective detection of issues \cite{24}. 
    \item \textbf{Activity and Updates:} This feature considers the events and activities relevant to the continued improvement and development of the testing tool. Testing tools should be updated as new vulnerabilities and attacks are discovered \cite{22, 86, 88}. This instills confidence in the capabilities of the testing tool to perform comprehensive assessment on the AI system and model. Additionally, an active community implies relevance and usefulness of the tool to its user base. Users may also request developers to incorporate features into the future releases of the tool.
    \item \textbf{Effectiveness:} This feature considers its accuracy in detecting threats rather than coverage. A tool with high false‑positive or false‑negative rates offers limited practical value, even if its other characteristics align well with the desired features defined above. Regardless, subjectivity in effectiveness evaluation remains a concern. To the best of our knowledge, currently there is no standardized benchmark for evaluating these tools, such as a benchmark set of vulnerable AI models. We attempted to address this through experiments on open-source tools against the same models. As a supplementary remark, following the “Standardization” desired feature, the tool’s benchmarking result should be measured with respect to a well-recognized existing taxonomy of threats or attacks from MITRE ATLAS \cite{56} or NIST AI 100-2E2025 \cite{29} to provide an objective evaluation of the tool, since different tools most often categorize their techniques differently.
\end{itemize}

\section{Metrics}

In this section, we examine how the proposed desirable features relate to the various metrics used in existing guidelines for comparing red‑teaming tools, as well as the effectiveness of each criterion. Table \ref{tab1} summarizes the metrics from a few guidelines and the relevant desired feature.
\small
\setlength{\tabcolsep}{5pt}
\begin{longtable}{@{\extracolsep{\fill}} p{0.24\linewidth} p{0.34\linewidth} p{0.34\linewidth}}
\caption{Mapping of metrics to desired features}\label{tab:mapping}\\
\toprule
Guidelines & Quantity & Relevant desired features \\
\midrule
\multirow[t]{3}{=}{Generative AI and Agentic Security Solutions Landscape (OWASP) \cite{90}} 
 & Type (Open Source / Proprietary) & Accessibility and Transparency \\ \cmidrule{2-3}
 & Gen AI / LLMSecOps & Models, modality and threat coverage \\ \cmidrule{2-3}
 & Top 10 for LLM Risk Coverage & Models, modality and threat coverage \\
\midrule
\multirow[t]{1}{=}{GenAI Red Teaming Guide (OWASP) \cite{86}} & Licensing Approach & Accessibility and Transparency \\
 & & \\
\midrule
\multirow[t]{10}{=}{Insights and Current Gaps in Open-Source LLM Vulnerability Scanners: A Comparative Analysis \cite{63}} & Attack Type & Models, modality and threat coverage \\ \cmidrule{2-3}
 & Guardrails Interface Support & N/A \\ \cmidrule{2-3}
 & Multi-language Support & Models, modality and threat coverage \\ \cmidrule{2-3}
 & Customizable Attacks Support & N/A \\ \cmidrule{2-3}
 & Automated Customization Support & N/A \\ \cmidrule{2-3}
 & Evaluation Result Explainability & Evidence Gathering and Documentation \\ \cmidrule{2-3}
 & Insecure Coding Tests Support & Models, modality and threat coverage \\ \cmidrule{2-3}
 & Volume of Attack Instances & Effectiveness \\ \cmidrule{2-3}
 & Attack Success Rate & Effectiveness \\ \cmidrule{2-3}
 & Margin of Error & Effectiveness \\
\midrule
\multirow[t]{18}{=}{AI security testing (OWASP AI Exchange) \cite{91}} & Licensing & Accessibility and Transparency \\ \cmidrule{2-3}
 & Providing Mitigation & N/A \\ \cmidrule{2-3}
 & API Availability & Automation and Integration \\ \cmidrule{2-3}
 & GitHub Stars & Activity and Updates \\ \cmidrule{2-3}
 & GitHub Forks & Activity and Updates \\ \cmidrule{2-3}
 & Number of Issues & Activity and Updates \\ \cmidrule{2-3}
 & Trend & Activity and Updates \\ \cmidrule{2-3}
 & Active Issues & Activity and Updates \\ \cmidrule{2-3}
 & Documentation & Activity and Updates \\ \cmidrule{2-3}
 & Discussion Forums & Activity and Updates \\ \cmidrule{2-3}
 & GitHub Contributors & Activity and Updates \\ \cmidrule{2-3}
 & Framework Support & Automation and Integration \\ \cmidrule{2-3}
 & Large-Scale Deployment & Automation and Integration \\ \cmidrule{2-3}
 & API Availability & Automation and Integration \\ \cmidrule{2-3}
 & Data Modality & Models, modality and threat coverage \\ \cmidrule{2-3}
 & Machine Learning Tasks & Models, modality and threat coverage \\ \cmidrule{2-3}
 & Framework Applicability & Automation and Integration \\ \cmidrule{2-3}
 & OWASP AI Exchange Threat Coverage & Models, modality and threat coverage \\
\midrule
\multirow[t]{3}{=}{Companion Guide on Securing AI System (CSA) \cite{24}} & License Type & Accessibility and Transparency \\
 & Model Type & Models, modality and threat coverage \\ \cmidrule{2-3}
 & Various from Pros \& Cons & Models, modality and threat coverage / \newline Activity and Updates / \newline Automation and Integration \\
\bottomrule
\label{tab1}
\end{longtable}

\subsection{Accessibility and Transparency}
In general, since licensing information is usually publicly available, this feature is straightforward to assess. CSA's Companion Guide on Securing AI Systems \cite{24} classified all non-proprietary tools simply as open source, while OWASP guidelines \cite{86,90,91} provide a more granular distinction between licensing types, such as MIT or Apache 2.0. 

\subsection{Standardization}
To the best of our knowledge, existing red teaming tool comparison guidelines do not evaluate a tool’s support for standards, although alignment to standards is recommended across various guidelines and regulations. This may be because AI red‑teaming is still an emerging area; development efforts tend to prioritize core functionality, such as effectiveness and coverage, over support for standardization. Moreover, many relevant standards are not yet widely adopted; as a result, integrating them into tools may not currently provide sufficient practical value. Few tools provide such support at this stage; hence it is not yet meaningful to use this as a criterion for comparison.

\subsection{Evidence Gathering and Documentation}
Evidence Gathering and Documentation is rarely used as a comparison criterion in existing guidelines. Only Brokman et al. \cite{63} considers a tool’s ability to explain its test case verdicts such as why an attack was deemed successful. Even then, it focuses solely on verdict explainability without considering other relevant evidence such as timestamps or test case inputs. However, most tools evaluated in the later section provided supporting information. This suggests that guidelines omit these aspects simply because they are already commonly provided and thus offer little value for distinguishing between tools. By contrast, the comprehensiveness and readability of an assessment report are hard to evaluate objectively or systematically, making these aspects best left for end users to assess directly.

\subsection{Automation and Integration}
This feature is assessed through multiple metrics across different guidelines. For example, OWASP's AI security testing \cite{91} evaluates a tool’s API availability, its compatibility with major frameworks such as TensorFlow or PyTorch, and its ability to work with different model families such as OpenAI or NVIDIA models. It also considers the tool’s scalability, i.e., whether it can support large scale industry deployments or is limited to research environments. This scalability criterion indirectly reflects the tool’s automation capabilities, since tools lacking robust automation are unlikely to be deployable at scale. CSA's Companion Guide on Securing AI Systems \cite{24} also assesses the tool’s integration capabilities with machine learning framework. Even though the guidelines do not explicitly assess a tool’s capability to conduct automated assessments, this can often be inferred from its integration features. While open‑source tools rarely include built‑in automation, having a CLI interface is usually sufficient for integration into existing CI/CD pipelines for automated assessments.

\subsection{Models, Modality, and Threat Coverage}
This feature is among those most emphasized by existing guidelines in tool comparisons. OWASP's Generative AI and Agentic Security Solutions Landscape \cite{90} maps tools to lifecycle operations using its LLMSecOps taxonomy (e.g., Agent Scanning, Sandboxed Testing) and coverage of the Top 10 LLM Risks. \cite{91} evaluates supported target modalities (e.g., text, image), machine learning tasks (e.g., object detection), and coverage of a different threat taxonomy. CSA's Companion Guide on Securing AI Systems \cite{24} focuses on supported model types such as LLMs or CNNs. Brokman et al. \cite{63} reviews supported attack types and whether tools handle multi language inputs or coding tests. It is evident that the effectiveness of the metrics depends on the quality of the taxonomy it is based on, as a poorly defined taxonomy renders the metrics ambiguous and ineffective. Furthermore, because each guideline employs its own threat taxonomy - with overlapping, merged, or slightly differing definitions - there is a clear need for a widely adopted and commonly agreed upon taxonomy. 

\subsection{Assessment Environments}
None of the guidelines examined evaluated how a tool can be deployed. That said, this aspect is generally straightforward to assess. Open-source tools are typically deployable on-premises, while deployment information is typically available on the vendor’s website through their pricing plans, or can be obtained by contacting the vendor directly if not. Another aspect the end user needs to evaluate is whether the tool operates in a white box or black box manner, which typically requires reviewing the tool’s documentation. In some cases, it may be useful to further break down the specific information about the target model that the tool can access, including the model’s weights, the LLM’s token level output probabilities, or simply its generated text, when assessing its capabilities.

\subsection{Activity and Updates}
This feature receives only moderate attention across the guidelines. In CSA's Companion Guide on Securing AI Systems \cite{24}, it is not assessed through a formal metric but rather is mentioned only as a drawback of tools that show low development activity. In contrast, OWASP's AI security testing \cite{91} provides a more detailed evaluation using several GitHub-based indicators, such as stars, number of active issues, and number of contributors. While these metrics serve as useful signals of an open-source project’s popularity and development activity, they can, in rare cases, be artificially inflated through bots. Moreover, GitHub metrics apply only to open-source tools and not commercial products, although some vendors do offer an open-source edition. For commercial tools, alternative indicators can include vendor revenue or release/update frequency and may be used to assess activity and adoption. It is also important to note that it is neither feasible nor meaningful to apply a single unified metric to compare open source and commercial tools such as release frequency, as they operate under fundamentally different business models.

\subsection{Effectiveness}
Although effectiveness is as critical as any other feature since an ineffective tool remains ineffective regardless of its integrability or threat coverage, only Brokman et al. \cite{63} attempts to evaluate it, and only within the context of LLM red teaming. This is likely due to the difficulty of measuring effectiveness objectively and consistently with minimal manual effort. In Brokman et al. \cite{63}, effectiveness is assessed using three metrics: volume of attack instances referring to the number of attack prompts, attack success rate reflecting the proportion that successfully compromise the model, and a manually estimated margin of error that accounts for misclassifications by automated verdicts, which require human review to estimate. 

However, these metrics are inadequate for evaluating a tool’s true effectiveness. First, the volume of attack instances does not reliably indicate diversity as many prompts may be highly similar. Beyond a certain threshold, additional samples merely yield diminishing returns and consume unnecessary resources. Furthermore, volume is not well-defined for multiturn attacks, or LLM generated prompts, where the number of queries is configurable. Second, attack success rate is heavily dependent on the dataset and prompts generated at runtime. A tool that repeatedly exploits the same vulnerability will attain a high success rate. What end users care about instead is the range of distinct attack approaches the tool can successfully execute. Thus, success rate should be measured per technique category while referring to a taxonomy \cite{29} and not per test case. Finally, estimating margin of error demands substantial manual effort and remains subjective and error prone. A panel of LLM judges could replace manual evaluation, which should be feasible given the simplicity of the judgment task and the capability of modern LLMs. 

\section{Experimental Setup}

We examined the proposed desired features used to evaluate red-teaming tools, and its effective it is in comparing tools.

\subsection{Models}

We employed the following open-source models from Ollama’s library \cite{92} for our experiments. We did not quantize any of the models for the experiments.
\begin{itemize}
    \item Llama 4-Scout-17B-16E-Instruct \cite{93}: Llama 4 Scout is a model with 17 billion active parameters and 16 experts. It is a multimodal model and is more powerful compared to previous Llama models, with a context window of 10M.
    \item Llama 3.2-3B-Instruct \cite{94}: It supports a context length of 128K tokens and is fine-tuned for use cases like assistant-like chat and agentic applications like knowledge retrieval and summarization. 
    \item Llama 2-7B-Chat \cite{95}: This model is created through supervised fine-tuning of pretrained Llama 2 for chat use cases. It has a context length of 4096 tokens and is fine-tuned on over 1 million human annotations.
    \item Vicuna v1.5 \cite{96}: Vicuna is a chat assistant trained by fine-tuning Llama 2 on user-shared conversations collected from ShareGPT. We used the 13B version which has a context length of 4096 tokens.
    \item DeepSeek v2-16B-Lite-Chat \cite{97}: DeepSeek-V2 is a Mixture-of-Experts (MoE) language model characterized by efficient training and interference. The Lite-Chat version has a context length of 32k tokens and is fine-tuned for chat use cases.
    \item Mistral 7B-Instruct-v0.3 \cite{98}: It is an instruction fine-tuned version of Mistral-7B-v0.3. It supports a context window of approximately 32k tokens.
\end{itemize}
    
\subsection{AI Security Testing Tools}
The following open-source tools were used for our experiments.
\begin{itemize}
    \item AgentFence \cite{61}: AgentFence is an open-source AI security testing framework that detects vulnerabilities in AI agents. It enables testing against predefined security probes and covers vulnerabilities such as prompt injection, role confusion, and system instruction leakage. It supports popular AI SDKs such as LangChain and OpenAI. Version 0.1.0 was used for experiments. 
    \item Garak \cite{33}: Designed by NVIDIA, Garak performs red teaming and evaluation of GenAI systems. It aims to determine whether a language model can be misled into producing undesired outcomes. Garak systematically probes for vulnerabilities such as hallucinations, data leaks, prompt injections, misinformation, generation of toxic content, jailbreaks, and a variety of other weaknesses. It assesses the model using a multitude of plugins and hundreds of probing techniques. Version 0.12.0 was used for experiments. 
    \item Giskard \cite{58}:  Giskard is an open-source library written in Python that detects performance, bias, and security issues in AI applications. The library covers RAG agents as well as traditional ML models for tabular data. It flags issues such as harmful content generation, hallucination, prompt injections, and sensitive information disclosure. Version 2.17.0 was used for experiments. 
    \item Promptfoo \cite{59}: Promptfoo is an open-source tool for automating LLM evaluation. It is flexibly designed to work with any LLM provider. It can speed up evaluations with caching, concurrency, and helps score outputs automatically. Version 0.117.5 was used for experiments. 
    \item Prompt Fuzzer \cite{60}: Developed by Prompt Security, Prompt Fuzzer evaluates the robustness of system prompts used in GenAI applications. It includes LLM-based attack simulations and provides a playground chat interface which enables users to repeatedly improve their system prompt and test it against a variety of techniques. Version 2.0.0 was used for experiments. 
    \item Whistleblower from Repello AI aims to infer the system prompt of a GenAI model based on its generated text outputs. The tool is designed based on the work of Zhang et al. \cite{99} but differs from it by leveraging in-context learning on a pre-trained LLM for predicting system prompts. Version 1.1.1 was used for experiments.
\end{itemize}

\section{Experimental results and analysis}

\subsection{Accessibility and Transparency}

All six tools—AgentFence, Garak, Giskard, Promptfoo, Prompt Fuzzer, and Whistleblower—demonstrate high Accessibility and Transparency, as each provides open‑source codebases and detailed documentation that allow adopters to fully inspect their internal mechanisms. This collective openness ensures that organizations can confidently evaluate the AI security testing tools operations before integrating them into their security workflows. 

\subsection{Standardization}

Four of the six tools tested do not explicitly align themselves with formal security standards for RMF, offering limited or no documented integration with frameworks such as NIST RMF \cite{22}, OWASP \cite{57,65,66,67,71,85,86,90,91}, MITRE ATLAS \cite{56}, or the EU AI Act \cite{68}. In contrast, Promptfoo stands out as the only tool that incorporates multiple industry taxonomies and regulatory frameworks into its testing methodology, indicating a significantly higher degree of standardization support than the others. On the other hand, although Giskard does not provide standardization with formal security standards for RMF, their scan reports indicate the AVID taxonomy categories of the detected vulnerabilities that are standardized for the ease of sharing.

\subsection{Evidence Gathering and Documentation}

All six tools generate detailed test reports containing prompt‑response pairs, providing organizations with concrete evidence of system behavior during adversarial testing. This level of documentation strengthens traceability across the AI lifecycle and gives stakeholders clearer visibility into vulnerabilities, making it easier to satisfy regulatory expectations for evidence collection and continuous oversight. However, the format of the report is not standardized. For example, AgentFence reports are via command line interface, while Garak reports are in JSONL and HTML format.

\subsection{Automation and Integration}

Table \ref{tab2} and \ref{tab3} show that integration capabilities vary widely across the tools: AgentFence, Garak, Giskard, and Whistleblower support meaningful integration with external model providers or endpoints, enabling more automated and repeatable assessment workflows. Promptfoo and Prompt Fuzzer lack such integration, making their assessment processes less automated and less consistent across runs. Overall, tools with broader integration support are better positioned to deliver scalable automation and smoother incorporation into an organization’s AI SDLC.

\begin{table*}
\caption{Analysis of open-source tools with respect to desired features}
\small
\setlength{\tabcolsep}{4pt}
\begin{tabular*}{\linewidth}{@{\extracolsep{\fill}} p{0.16\linewidth} p{0.28\linewidth} p{0.22\linewidth} p{0.28\linewidth}}
\toprule
Tool 
& Accessibility and Transparency 
& Standardization 
& Evidence Gathering and Documentation \\
\midrule
AgentFence 
& High (open source + detailed documentation). Moderate for commercial version (has closed source features) 
& N/A 
& Generate detailed test report with prompt-response pairs \\
\midrule
Garak 
& High (open source + detailed documentation) 
& N/A 
& Generate detailed test report with prompt-response pairs \\
\midrule
Giskard 
& High (open source + detailed documentation). Moderate for enterprise hub (closed source) 
& AI Vulnerable Database AVID 
& Generate detailed test report with prompt-response pairs \\
\midrule
Promptfoo 
& High (open source + detailed documentation) 
& MITRE ATLAS, NIST AI RMF, OWASP, EU AI Act 
& Generate detailed test report with prompt-response pairs \\
\midrule
Prompt Fuzzer 
& High (open source + detailed documentation) 
& N/A 
& Generate detailed test report with prompt-response pairs \\
\midrule
Whistle-Blower 
& High (open source + detailed documentation) 
& N/A 
& Generate detailed test report with prompt-response pairs \\
\bottomrule
\end{tabular*}
\label{tab2}
\end{table*}

\begin{table*}
\caption{Analysis of open-source tools with respect to desired features (cont.)}
\small
\setlength{\tabcolsep}{4pt}
\begin{tabular*}{\linewidth}{@{\extracolsep{\fill}} p{0.18\linewidth} p{0.42\linewidth} p{0.32\linewidth}}
\toprule
Tool 
& Automation \& Integration 
& Activity \& Update \\
\midrule
AgentFence 
& Can integrate with Hugging Face, Ollama, LangChain, OpenAI (for judge) 
& 51 stars on Github, last updated in March 2025 \\
\midrule
Garak 
& Wide range of integration: Cohere (API), Ggml (Local), Hugging Face (API/Local/Private Endpoint), OpenAI (API), Replicate (API / Private Endpoint), Function (Local: self-model), etc. 
& 7.1k stars on Github, is still having new commit every week \\
\midrule
Giskard 
& Can generate NeMo Guardrails. The assessment process is repeatable 
& 5.1k stars on Github, is still having new commit every week \\
\midrule
Promptfoo 
& Has no integration. The assessment process is not repeatable 
& 10.6k stars on Github, is still having new commit every week \\
\midrule
Prompt Fuzzer 
& Has no integration. The assessment process is not repeatable 
& 636 stars on Github, is still having new commit every week \\
\midrule
Whistle-Blower 
& Can integrate with Hugging Face, Ollama, LangChain, OpenAI (for judge) 
& 151 stars on Github, last updated in October 2025 \\
\bottomrule
\end{tabular*}
\label{tab3}
\end{table*}

\subsection{Activity and update}

Four of the six tools exhibit strong ongoing development, with Garak, Giskard, Promptfoo, and Prompt Fuzzer receiving regular weekly commits, indicating active maintenance, rapid response to newly discovered vulnerabilities, and a healthy contributor ecosystem. AgentFence and WhistleBlower, while still updated, show less frequent activity, suggesting slower update cycles that may limit their responsiveness to emerging threats compared to the more actively maintained tools.

\subsection{Effectiveness}

As observed, AI red-teaming tools available employ their own system for categorizing the attack types they support. The lack of consistency makes it challenging to compare tools directly. To address this, we propose the creation of a standardized table that records, for each tool, each model, and each attack type grounded by a trustworthy taxonomy, whether the tool can successfully conduct that attack type on the given model.

To initiate this effort, we map the attack type of the six red teaming tools to the NIST AI 100-2 E2025 report \cite{29}, which is comprehensive, rigorously peer-reviewed, and widely recognized as a standard reference in the field. Its structured definitions ensure consistency across evaluations and provide a credible foundation for comparing tools in a reproducible manner. After aligning the tools within the NIST report, we observe a predominantly common set of targeted vulnerabilities: prompt injection (NISTAML.018), prompt extraction (NISTAML.035), privacy compromises (NISTAML.03), misuse violations (NISTAML.04), data extraction (NISTAML.038), and misaligned outputs (NISTAML.027). The effectiveness of each tool for each vulnerability is shown in Table \ref{tab4}. This standardization enables more meaningful comparisons across tools. For instance, although both Prompt Fuzzer and Whistleblower include techniques that map to Prompt Extraction (NISTAML.035), Prompt Fuzzer demonstrates significantly greater effectiveness, identifying vulnerabilities in all six models, whereas Whistleblower shows less effectiveness.

\begin{table*}
\caption{Analysis of open-source tools with respect to Effectiveness}
\small
\setlength{\tabcolsep}{4pt}
\begin{tabular*}{\linewidth}{@{\extracolsep{\fill}} p{0.2\linewidth} p{0.1\linewidth} p{0.1\linewidth} p{0.1\linewidth} p{0.1\linewidth} p{0.1\linewidth} p{0.1\linewidth}}
\toprule
Tool & Agent Fence & Garak & Giskard & Promptfoo & Prompt Fuzzer & Whistle blower \\
\midrule
Prompt Injection (NISTAML.018) & 4 & 5 & 6 & - & 6 & - \\
\midrule
Prompt extraction (NISTAML.035) & 2 & - & - & - & 6 & 0 \\
\midrule
Misuse Violations (NISTAML.04) & 0 & 5 & - & 6 & - & - \\
\midrule
Data extraction (NISTAML.038) & 5 & 5 & - & 6 & - & - \\
\midrule
Misaligned Outputs (NISTAML.027) & - & - & 5 & 6 & 5 & - \\
\midrule
Privacy Compromises (NISTAML.03) & - & - & 5 & 6 & - & - \\
\midrule
Compromising connected resources (NISTAML.039) & - & - & - & 6 & - & - \\
\midrule
Indirect prompt injection (NISTAML.015) & - & - & - & 6 & - & - \\
\bottomrule
\end{tabular*}
\label{tab4}
\end{table*}

\section{Conclusion}

AI offers many benefits to organizations, helping to automate decision-making and enhance the efficiency of tasks. The increased adoption of AI also brought concerns of its security as this opens additional attack surfaces. The rise of threats pertaining to AI led to the publication of frameworks and guidelines for the secure development and usage of AI systems. In this study, we examined multiple guidelines to derive eight desirable features that organizations should prioritize when evaluating third‑party AI red‑teaming tools. We mapped these features to existing comparison metrics to assess their practical applicability and conduct experiments on popular open-source LLM red‑teaming frameworks to demonstrate how the proposed features can be utilized. This work hopes to assist organizations in selecting the appropriate AI security testing tool for secure and safe development of AI systems and models. 

As future work, the effectiveness, and capabilities of AI defensive security mechanisms can be evaluated to protect AI systems and models against any vulnerabilities and weaknesses. These include defensive techniques and controls such as guardrails that monitor, detect, and mitigate against potentially harmful inputs and outputs from the AI system and model. They can be applied during the preprocessing and postprocessing stages, such as adversarial training, and leverage transformers. This on-going process of identifying and addressing vulnerabilities and weaknesses as organizations attempt to secure and protect their AI systems and models against the ever-evolving threat landscape throughout the AI SDL.

\bibliography{submission-ref}

\end{document}